\documentclass[aps,prd,twocolumn,showpacs,groupedaddress]{revtex4}
\usepackage{epsfig}
\usepackage{longtable}
\usepackage{amsfonts}

\newcommand{\tr}{{\rm tr}}
\newcommand{\ot}{\omega_t}
\newcommand{\oo}{{\omega_\oplus}}
\newcommand{\To}{T_\oplus}

\newcommand{\om}{\omega}

\begin{document}

\title{Tests of Lorentz invariance using hydrogen molecules}
\author{Holger M\"{u}ller}
\altaffiliation{Now at Physics Department, Varian Bldg., Room 226, Stanford University,
Stanford, CA 94305-4060, Phone: (650) 725-2354, Fax: (650) 723-9173}
\email{holgerm@stanford.edu}
\author{Sven Herrmann}
\author{Alejandro Saenz}
\author{Achim Peters}
\affiliation{Institut f\"{u}r Physik, Humboldt-Universit\"{a}t zu Berlin, Hausvogteiplatz 5-7, 10117 Berlin, Germany.\\
Tel. +49 (30) 2093-4907, Fax +49 (30) 2093-4718}
\email{achim.peters@physik.hu-berlin.de}
\author{Claus L\"{a}mmerzahl}
\affiliation{ZARM (Center for Applied Space Technology and Microgravity), Universit\"{a}t Bremen,
Am Fallturm, 28359 Bremen, Germany}

\date{}

\begin{abstract}
We discuss the consequences of Lorentz violation (as expressed
within the Lorentz-violating extension of the standard model) for
the hydrogen molecule, which represents a generic model of a
molecular binding. Lorentz-violating shifts of electronic,
vibrational and rotational energy levels, and of the internuclear
distance are calculated. This offers the possibility of obtaining
improved bounds on Lorentz invariance by experiments using
molecules.
\end{abstract}


\maketitle

\section{Introduction}
A violation of Lorentz invariance occurs in many current models of
quantum gravity. While such theories typically operate on the Planck energy scale, low-energy remnants of Planck scale physics might break Lorentz invariance in the equations of motion of some or all of the particles of the standard model. Such Lorentz violation is generally described by the standard model extension (SME) \cite{KostSME}. Currently, there is a large experimental effort to find sharp limits on Lorentz violation for the different sectors of the standard model, and many types of Lorentz violation for different particles have been constrained \cite{Kos03,KL99}.

Here, we consider the quantum electrodynamical sector of the SME that includes photons and electrons. Stringent limits on Lorentz violation in the other sectors allow us to neglect the possibility of Lorentz violation in other sectors for this work. For photons, atoms (e.g., also
for atomic hydrogen \cite{Blu99}), and many sub-atomic particles, consequences of Lorentz violation have been studied extensively, see \cite{Kos03,KL99,Blu99,SPS} and references therein. However, many high-precision experiments are performed on chemically bonded systems like molecules (e.g., high resolution spectroscopy) or solids (e.g. experiments with macroscopic cavity resonators \cite{BrilletHall,MHBSP,MMalii,Braxmaier}). A study of the influence of Lorentz violation on chemical bonds is thus interesting.

As a generic model of chemical bonds, we investigate neutral and
ionized molecular hydrogen H$_2$ and H$_2^+$ under the influence
of Lorentz violation in the electrons' equation of motion. These
molecules are simple enough so that a specific wave--function can
be used as an ansatz for the calculations. At the same time, they
exhibit interesting features not found in simpler systems (like
hydrogen atoms): For example, a preferred direction is singled out
by the molecules' axis; the internuclear distance defines a length
standard whose behavior in the case of Lorentz violation can
explicitly be studied (and compared to other different length
standards, like crystals \cite{resSME,TTSME} or the distance
traveled by a ray of light within a certain time).

Experiments aiming for sympathetic cooling (to mK temperature) and
high-resolution spectroscopy of H$_2^+$ and HD$^+$ are currently
under way \cite{Schiller03,Froe04}. Using the theory presented
here, limits on Lorentz violation can be obtained in these
experiments by searching for a Lorentz-violating shift in the
transition frequencies. With the high resolution that is possible
in frequency metrology and the suppression of line broadening
mechanisms for the cooled molecules, such experiments may improve
the present bounds on particular Lorentz violating parameters for
the electron.

We treat the hydrogen molecule using the Born-Oppenheimer model
that is described in textbooks, e.g. \cite{fluegge}. The motion of
the nuclei is neglected and approximate electron wave functions
are obtained by linear combination of atomic orbitals. The
Born-Oppenheimer model is a basic model with sufficient precision
for our purposes, as we do not need to predict the absolute values
of the quantities, but only the Lorentz violating shifts of them.
We work to first order in the Lorentz violating parameters
throughout. Since the nuclei are assumed to be point-like, our
results are valid for H$_2$ and H$_2^+$ as well as for HD and
HD$^+$.

In Sec. \ref{Hydrogencalculations}, we calculate the matrix elements of the Lorentz violating quantities from the SME for H$_2$ and H$_2^+$, taking into account that usually the molecules will be in an angular momentum eigenstate. In Sec. \ref{modprop}, we calculate the modified ground state energy, bond length, and frequencies of rotational and vibrational transitions. In Sec. \ref{tests}, we discuss the possibility of improving present limits on Lorentz violation by experiments using molecules. The appendix gives the explicit time-dependence of the hypothetical signal for Lorentz violation.

\subsection*{Standard model extension (SME)}

The SME \cite{KostSME,KL99,Blu99,Kos03} starts from a Lagrangian
formulation of the standard model, adding all possible observer
Lorentz scalars that can be formed from the known particles and
Lorentz tensors. The non-relativistic Schr\"{o}dinger Hamiltonian
$h=\hat{h}+\delta h$ of a free electron derived from the SME
Lagrangian is the sum of the usual free-particle Hamiltonian $\hat
h$ and a Lorentz-violating term. Disregarding a constant term that
has no physical consequences and terms proportional to odd powers
of the particle momentum that vanish in the center of mass system
\cite{KostSME,KL99,Kos99b,LaemmerzahlCQG},
\begin{equation}\label{reducedhamil}
\delta h = m_e c^2 B'_j \sigma^j + E_{jk} \frac{p_j p_k}{m_e} +
F_{jkl} \frac{p_j p_k}{m_e}\sigma^l\,.
\end{equation}
Here, $p_j$ ($j=1,2,3$) are the components of the 3-momentum,
$\sigma^j$ are the Pauli matrices, and $m$ is the mass of the
electron. The abbreviations
\begin{eqnarray}
B'_j & = & -\frac{b_j}{m} + d_{j0} - \frac12
\varepsilon_{jkl}g_{kl0} + \frac{1}{2m} \varepsilon_{jkl}H_{kl} \,
, \\ E'_{jk} & = &-c_{jk}-\frac{1}{2}c_{00}\delta_{jk} \, ,
\label{hamiltonianparameters} \\ F'_{jkl} & =
&\bigg[(d_{0j}+d_{j0}) -\frac12\bigg( \frac{b_j}{m}+d_{j0}+\frac12
\varepsilon_{jmn}g_{mn0} \nonumber\\ & &
+\frac{1}{2m}\varepsilon_{jmn}H_{mn} \bigg)\bigg]
\delta_{kl}+\frac 12\left(\frac{b_l}{m}+\frac
12\varepsilon_{lmn}g_{mn0}\right)\delta_{jk} \nonumber \\ & &
-\varepsilon_{jlm} (g_{m0k}+g_{mk0}) \, . 
\end{eqnarray}
contain the Lorentz tensors $b_\mu, c_{\mu \nu}, d_{\mu\nu},
g_{\lambda \mu \nu}$, and $H_{\mu \nu}$ that encode Lorentz
violation for the fermions within the SME defined in
\cite{KostSME,Kos03}. In this work, we deal with electrons, so all
these parameters are electron parameters. For neutral hydrogen
H$_2$, this perturbation has to be summed over the two electrons.
To first order in the changes, the influence of the Lorentz
violating perturbation is given by its matrix elements.

\section{Molecular hydrogen}\label{Hydrogencalculations}

We will first assume non-rotating molecules; the rotation of the molecules will be treated
subsequently. Practically, the influence of the rotation on the other properties of the molecule
is small, so we may neglect it here.

\subsection{Neutral hydrogen molecule}

\begin{figure}
\centering \epsfig{file=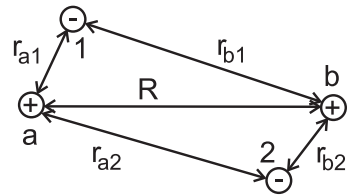, width=0.25\textwidth} \caption{\label{vectors} Definition of
coordinates. $a$ and $b$ denote the nuclei, $1$ and $2$ the electrons.}
\end{figure}

We are using atomic units $m=\hbar=e=1$, where $e$ is the electron's charge. We denote the two nuclei $a$ and $b$, and enumerate the electrons $1,2$. Let $(x_1)_j$ and $(x_2)_j$ be the spatial components of the position of electron 1 and 2, respectively, and $(p_1)_j$, $(p_2)_j$ those of their momenta. Let $\vec r_{a1}$ be
the distance between nucleus $a$ and electron 1, and so forth, and $\vec R$ the distance between the nuclei (see Fig. \ref{vectors}). For this calculation, we adopt coordinates such that $\vec R$ is parallel to the $z$ axis. 

The antisymmetric wave function of H$_2$ can be written as the product of a spatial function $\Psi_r$ and a spin function that is antisymmetric under exchange of the electrons. Thus, the matrix elements of the spin--dependent terms of $\delta h$ vanish and the H$_2$ molecule will only be sensitive to the Lorentz violating quantity $E_{jk}$ of Eq. (\ref{reducedhamil}).

We make the usual ansatz for the spatial wave function of H$_2$
\cite{fluegge}:
\begin{equation}\label{LCAO}
\Psi_r=\alpha [f(\vec r_{a1})f(\vec r_{b2})+f(\vec r_{b1})f(\vec r_{a2})]
\end{equation}
with a normalization factor $\alpha =1/\sqrt{2(1+S^2)}$, where
\begin{equation}\label{overlap}
S=\int f^*(\vec r_{a1})f(\vec r_{b1})d^3x_1= \left[1+\gamma R+ \frac 13 (\gamma
R)^2\right]e^{-\gamma R}
\end{equation}
is the overlap integral. The function
\begin{equation}\label{wellenfkt}
f(\vec r)=\sqrt{\frac{\gamma^3}{\pi}}e^{-\gamma r}
\end{equation}
satisfies $\int |f(\vec r)|^2 d^3 r = 1$. For $\gamma=1$, it is
the usual single-electron wave function of the hydrogen atom in
the ground state. The ground state of H$_2$ is found by minimizing
the energy $U_{\rm H_2}$ of the system as a function of $R$ and
$\gamma$. For H$_2$, the global minimum is obtained at $R_0=1.41$ with $\gamma=1.166$:
$U_{\rm H_2}(R_0)=-1.139$ \cite{fluegge,missprints}. For each $R$,
there is a $\gamma(R)$ that gives a local minimum for $U_{\rm
H_2}$. When calculating the dependence of $U_{\rm H_2}$ on $R$,
the dependence of $\gamma(R)$ has to be taken into account, since
the electronic wave functions will adjust themselves for each $R$
rapidly compared to, e.g., the molecular vibrations.

The Lorentz violating changes are induced by the matrix elements $\left<p_ip_j\right>$. Due to
symmetry under inversion of the $x$ and $y$ axes (that also holds for HD and HD$^+$), $\left<p_ip_j\right>=0$ unless $i=j$ (which can also be verified by explicit calculation
in analogy to Eq. (\ref{quadratures})). Furthermore $\left<p_x^2\right>=\left<p_y^2\right>$. Thus,
\begin{equation}
\left<\delta h \right>_{\rm H_2}= E_{ii} \left<p_i p_i
\right>_{\rm H_2}
\end{equation}
with
\begin{eqnarray}\label{deltah}
\left<p_i p_i \right>_{\rm H_2} & = & -\int \Psi_r^* [ (\partial_i^2)_1 + (\partial_i^2)_2 ]\Psi_r
d^3(x)_1 d^3(x)_2 \nonumber
\\ & = &  -4 \alpha^2 \left[\mathcal A'_i + S \mathcal A''_i
\right]
\end{eqnarray}
where we denoted $(\partial_j^2)_1 \equiv \partial^2/(\partial (x_1)^2)$, and
\begin{eqnarray}\label{ajstrichstrich}
\mathcal A'_j & = & \int f^*(\vec r_{a1})\frac{\partial^2}{\partial (x_1)_j^2}f(\vec r_{a1}) d^3(x)_1
= -\frac {\gamma^2}{3} \, , \nonumber \\ \mathcal A''_j & = & \int f(\vec r_{b1})
\frac{\partial^2}{\partial (x_1)_j^2}f^*(\vec r_{a1})d^3(x)_1 \, .
\end{eqnarray}
The first integral is evaluated using elementary integrations in polar coordinates with the
$\theta=0$ direction parallel to $x_j$. For $\mathcal A''_j$, we insert Eq. (\ref{wellenfkt}):
\begin{equation}\label{astrich}
\mathcal A''_j=-\frac{\gamma^4}{\pi} \int e^{-\gamma
(r_{a1}+r_{b1})}\left(\frac{1}{r_{a1}}-\frac{(x_j)_1^2}{r_{a1}^3}-\gamma
\frac{(x_j)_1^2}{r_{a1}^2} \right)d^3 x_1 \, .
\end{equation}
We now use polar coordinates $(r,\theta,\phi)$ with nucleus $a$ as center and the $\theta=0$ axis
pointing towards nucleus $b$:
\begin{equation}\label{quadratures}
\mathcal A''_j=\frac{\gamma^4}{\pi} \int e^{-\gamma (r+r_{b1})}\left(\frac{(x^j)_1^2}{r}+\gamma
(x^j)_1^2-r \right)\sin\theta dr d\theta d\phi
\end{equation}
with
\begin{eqnarray}
x^1 \equiv x & = & r \cos \theta \, , \nonumber \\  x^2 \equiv y & = & r \sin \theta \sin \phi \, , \nonumber \\ x^3 \equiv z
& = &  r \sin \theta \cos \phi \, , \nonumber \\  r_{b1}^2  & = &  r^2-z^2+(z+R)^2 \, .
\end{eqnarray}
Numerical computation yields the values for $\left<p_x p_x \right>_{\rm H_2}=\left<p_y p_y
\right>_{\rm H_2}$, $\left<p_z p_z \right>_{\rm H_2}$ and their derivatives $d^n \left<p_i p_i
\right>_{\rm H_2}/ d R^n$ given in Tab. \ref{integrals}.

\begin{table}[t]
\centering \caption{\label{integrals} $\left<p_x p_x \right>=\left<p_y p_y \right>$,
$\left<p_z p_z \right>$ and their derivatives $(d^n \left<p_i p_i \right>/ d R^n)_{R_0}$ for H$_2$ (at $R_0=1.414$) and H$_2^+$ ($R_0=2.00$). Also shown is the unperturbed energy $U(R)$ and its derivatives $(d^nU/dR^n)_{R_0}$. All these are total derivatives, with $\gamma(R)$ (also tabulated together with its derivatives) also depending on $R$. The first derivative
$(dU/dR)_{R_0}$ vanishes, since $R_0$ minimizes $U$.}
\begin{center}
\begin{tabular}{ccccc}
\hline $n$ & 0 & 1 & 2 & 3 \\ \hline $\left<p_x p_x \right>_{\rm H_2}$ & 0.833 & -0.344 & 0.454 &
\\ $\left<p_z p_z \right>_{\rm H_2}$ & 0.612 & -0.266 &
0.501 &  \\ $\gamma_{\rm H_2}$ & 1.166 & -0.238 & 0.233 & -0.150 \\ $U_{\rm H_2}$ & -1.139 & 0.000
& 0.374 & -0.849 \\ $\left< p_x p_x \right>_{\rm H_2^+} $ & 0.451 & -0.147 & 0.147 &
\\ $\left< p_z p_z \right>_{\rm H_2^+} $ & 0.271 & -0.084 & 0.134 &
\\ $\gamma_{\rm H_2^+}$ & 1.238 & -0.203 & 0.150 & -0.112 \\ $U_{\rm H_2^+}$ & -0.587 & 0.000 & 0.100 & -0.259 \\ \hline
\end{tabular}
\end{center}
\end{table}

\subsection{Ionized hydrogen molecule}

Ionized hydrogen H$_2^+$ can be treated analogously. However, since there is only one electron,
the spin dependent terms do not vanish, and we have to deal with the full Lorentz violating
correction of the Hamiltonian, Eq. (\ref{reducedhamil}). The wave function for H$_2$ is modeled as
\begin{equation}
\psi=\beta [f(\vec r_{a})+f(\vec r_{b})]
\end{equation}
with a normalization factor $\beta=1/\sqrt{1+S}$ and $f$ as defined in Eq. (\ref{wellenfkt}). For
H$_2^+$, $\gamma=1.24$ and $R_0=2.00$ give minimum total energy \cite{fluegge}. Again, $\gamma
=\gamma(R)$ has to be taken into account. We can write the Lorentz violating contribution as
\begin{equation}
\left<\delta h\right>_{\rm H_2^+} = B_i \sigma^i - \tilde E_{ii}
\left< p_i p_i \right>_{\rm H_2^+}
\end{equation}
(again, $\left<p_ip_j\right>=0$ for $i\neq j$) with
\begin{equation}
\tilde E_{jk} = E_{jk}+F_{jkl} \sigma_l\, .
\end{equation}
and
\begin{equation}
\left< p_j p_j \right>_{\rm H_2^+} =\int \psi^* \frac{\partial^2}{\partial x_j^2} \psi d^3x = -
\beta (\mathcal A'_j+\mathcal A''_j) \, .
\end{equation}
With $\mathcal A'_j, \mathcal A''_j$ as above. The results are given in Tab. \ref{integrals}.

\subsection{Rotation of the molecules}

As shown above, the expectation value of the Lorentz violation for
neutral and ionized hydrogen depend only on two parameter
combinations that we may choose as the trace $\tr(E)$ and $E_3$,
\begin{eqnarray}
\tr(E)&=&E_{xx}+E_{yy}+E_{zz},\nonumber\\
E_3&=& E_{xx}+E_{yy}-2E_{zz}=\tr(E)-3E_{zz},
\end{eqnarray}
and similar for $\tilde E$. Spatial rotations, given by the matrix
\begin{equation}
R(\theta,\phi)=\left(\begin{array}{ccc} \cos\phi & \cos\theta \sin\phi & \sin\phi \sin\theta \\
-\sin\phi & \cos\phi \cos\theta & \cos\phi \sin\theta \\ 0 &  -\sin\theta &  \cos\theta
\end{array}\right) \, ,
\end{equation}
will leave tr$(E)$ unchanged, while $E_{zz} \rightarrow
E_{zz}(\theta,\phi)$ with
\begin{eqnarray}\label{deltarh2rotprae}
E_{zz}(\theta,\phi) = \sin^2\theta[E_{yy}\cos^2\phi +
E_{xx}\sin^2\phi
 +E_{xy}\sin2\phi] \nonumber
\\ +\sin2\theta[E_{yz}\cos\phi +E_{xz}\sin\phi] +E_{zz} \cos^2\theta \, , \quad
\end{eqnarray}
which leads to a change $E_3 \rightarrow E_3(\theta,\phi)$.

Now we consider molecules whose axes are not fixed: Due to rotation invariance without Lorentz
violation, the wave function of the unperturbed molecules can be written as the product of the
above wave functions and spherical harmonics $Y_l^m(\theta,\phi)$. Thus, $\delta h$ has to be
averaged with the angular wave functions $\left|lm\right>$. Due to the rotation invariance of
$\tr(E)$, $\left<lm|\tr(E)|lm\right>=\tr(E)$. For $(E_3)_l^m:=\left<lm|E_3|lm\right>$, we express
the sine and cosine functions in Eq. (\ref{deltarh2rotprae}) in terms of spherical harmonics:
\begin{eqnarray}\label{ezzylm}
E_3(\theta,\phi) =-\frac 13 \tr(E) +E_3\sqrt{\frac{4\pi}{5}}Y_2^0
+E_{xx}X_2^{2+}\nonumber
\\+E_{yy}X_2^{2+} +2iE_{xy}X_2^{2-} +2E_{yz}X_2^{1-}-2iE_{xz}X_2^{1+}
\end{eqnarray}
where we used the abbreviation
\begin{equation}
X_l^{m\pm}=\sqrt{\frac{6\pi}{5}}(Y_l^m\pm Y_l^{-m}) \, .
\end{equation}
The average is calculated by an application of the Wigner-Eckart theorem
\cite{Landau}. $\left<lm|Y_l^m|lm\right>=0$ unless $m=0$, which removes all but the first two terms
of Eq. (\ref{ezzylm}). After some algebra, we find
\begin{eqnarray}\label{ezzrot}
(E_3)_l^m &=& -(-1)^{m}(2l+1) \left(\begin{array}{ccc} l & l & 2
\\ 0 & 0 & 0
\end{array} \right)
 \left(\begin{array}{ccc} l & l & 2 \\ m & -m & 0 \end{array}
\right)E_3 \nonumber \\ &=&
2\frac{(l+1)[3m^2-l(l+1)]}{(2l-1)(2l+2)(2l+3)}E_3\quad
\end{eqnarray}
and analogous for $\tilde E_{zz}$. The brackets are the
$3j$-symbols \cite{Landau}.

The situation that the quantization axis itself is rotated (e.g.
by fixing it in the lab and using Earth's rotation) can be most
easily treated by transforming the Lorentz-violating SME tensor
$c_{\mu\nu}$ into a co-rotating frame, as dicussed in the
appendix. (The same situation can be described by a time-dependent
superposition of the eigenfunctions $Y_l^m$ for a fixed
quantization axis. Some elements of $E_{ij}$ would enter the
expectation value in terms of the form
$\left<l_1m_l|Y_l^m|l_2m_2\right>$, which is nonzero for
$m_2-m_1+m=0$.)


\section{Modified properties of the ground states}\label{modprop}

\subsection{Energy}

The energy $U$ of the ground state is shifted by the expectation
value $\left< \delta h \right>$. From the numerical values given in
table \ref{integrals}, we obtain for H$_2$ and $H_2^+$,
respectively \cite{remark}
\begin{eqnarray}\label{deltaU}
\left(\frac{\delta U}{U}\right)_{\rm H_2} & = & -0.667 \tr(E)
-0.065 (E_3)_l^m \, , \nonumber \\ \left(\frac{\delta
U}{U}\right)_{\rm H_2^+} & = & -0.667 \tr(\tilde E) -0.102 (\tilde
E_3)_l^m +\frac{m_ec^2}{U_{\rm H_2^+}}B_i\sigma^i  \, .\quad
\end{eqnarray}
Here, the molecules are assumed to be in an eigenstate of the
angular momentum with quantum numbers $l$ and $m$. For molecules
with a fixed orientation, $(E_3)_l^m$ is to be replaced by $E_3$.
For H$_2$ in the ground state, the spin--dependent term
proportional to $m_ec^2/U_{\rm H_2^+}\simeq 3.2\times 10^4$ drops
out, as the spins of the electrons are antiparallel.

Precision ab initio calculations \cite{calculations} and
measurements \cite{measurements,Gil92} of the ground state energy
of H$_2$ agree to an impressive precision of $\delta U/U\lesssim
1\times 10^{-7}$. According to Eq. (\ref{deltaU}), this gives an
upper limit of $|\tr(E)| \lesssim 1 \times 10^{-7}$ and
$E_3\lesssim 10^{-6}$ (as molecules with many different
orientations of the quantization axis and combinations of $l,m$
enter the measurements). However, more precise limits were already
derived \cite{resSME} from electromagnetic cavity
experiments \cite{BrilletHall,MHBSP}, as discussed below.

\subsection{Bond length}

The change of the bond length $R_0$ can be calculated by minimizing the total energy
\begin{equation}
U=U_0+\left<(\delta h)_{\rm H_2}\right>
\end{equation}
of the molecule. $U_0$ is the energy without Lorentz violating terms, as it is calculated in
the literature. It has a minimum at $R_0$, the bond length of H$_2$. With Lorentz violating terms,
\begin{equation}
U(R_0+\delta R) = \mbox{const.}+\frac 12 \frac{\partial^2 U_0}{\partial R^2} (\delta R)^2+ \frac{\partial
\left<(\delta h)\right>}{\partial R} \delta R \, ,
\end{equation}
so the new energy minimum will be at a modified length $R=R_0 + (\delta R)_{\rm min}$. Setting
$\partial U(R)/\partial R = 0$ and inserting the numerical values from Tab. \ref{integrals} leads
to
\begin{eqnarray}\label{deltarh2rot}
\left(\frac{\delta R}{R_0}\right)_{\rm H_2} & = & 0.603 \tr(E)
+0.050 (E_3)_l^m \, , \nonumber \\ \left(\frac{\delta
R}{R_0}\right)_{\rm H_2^+} & = & 0.632 \tr(\tilde E)+0.106 (\tilde
E_3)_l^m \, .
\end{eqnarray}
The term proportional to $B_j$ does not contribute here and below, as it does not depend on the
internuclear distance $R$. As the bond length has not been measured to a similar precision as
energy levels, a comparison between theory and experiment does presently not lead to interesting
limits on Lorentz violation.

\subsection{Vibrational transitions}

The energy levels of a quantized harmonic oscillator $U_v = \left(v + 1/2 \right) \hbar \omega$,
where $\omega$ is the resonance frequency and $v = 1,2, \ldots$ the vibrational quantum number, are equidistant with a difference $\hbar \omega$ between neighbors. For the hydrogen molecule, $\omega=
\sqrt{k/\bar m}$, where $k=\partial^2 U/\partial R^2$ and $\bar m=m_am_b/(m_a+m_b)$ is
the reduced mass of the nuclei. For the change of $\partial^2 U/\partial R^2$, we take into account both the addition of $\delta h$ as well as the change due to the shift $\delta R$ in the internuclear distance:
\begin{equation}
\left(\frac{\partial^2 U}{\partial R^2}\right)_{R_0+\delta R}=\left(\frac{\partial^2 U}{\partial R^2}\right)_{R_0}+\left( \frac{\partial \left<\delta h\right>}{\partial R^2}\right)_{R_0} + \left(\frac{\partial^3 U}{\partial R^3}\right)_{R_0} \delta R \, .
\end{equation}
(We can use $\left< \delta h \right>(R_0+\delta R) \approx \left< \delta h \right>(R_0)$, since the
difference is of second order in the Lorentz violating terms.) Since $\omega$ is proportional to
$\sqrt{k}$,
\begin{eqnarray}\label{deltak}
\frac{\delta \omega}{\omega}=\frac 12\frac{\delta k}{k}  =\frac{1}{2k}\left[\left( \frac{\partial^3
U}{\partial R^3}\right)_{R_0} \delta R + \left(\frac{\partial^2 \left< \delta h \right>}{\partial R^2}\right)_{R_0}\right]
\, .
\end{eqnarray}
Inserting the numerical values given in Tab. \ref{integrals}, we obtain
\begin{eqnarray}\label{deltaomvib}
\left(\frac{\delta \omega}{\omega}\right)_{\rm H_2} & = &
-0.337\tr(E)-0.100(E_3)_l^m \, , \nonumber
\\ \left(\frac{\delta \omega}{\omega}\right)_{\rm H_2^+} & = & -0.926\tr(\tilde E)-0.254(\tilde E_3)_l^m \, .
\end{eqnarray}

The above treatement of the vibrational transitions is based on a harmonic approximation for the core-core potential near equilibrium. This approximation is relatively accurate for low excitations. For example, the energy of the $v=4$ vibrational level of $H_2^+$ within the harmonic approximation differs from the realistic value by about 0.6\%. Thus, for the vibrational transitions used in the experiments to be discussed below, the harmonic approximation is sufficient.

\subsection{Rotational transitions}

The rotation of the molecule without Lorentz violation is
characterized by an energy $H_{\rm rot}=1/(2\bar m R_0^2)\vec L^2$,
where $\vec L$ is the angular momentum operator with eigenvalues
$l(l+1)$ ($l=0,1,2,3, \ldots$). Transitions between the rotational
energy levels thus have a frequency of
\begin{equation}
\omega_{\rm rot}(l\rightarrow l')=\frac{1}{2\hbar \bar m
R_0^2}\left(\frac{1}{l(l+1)}-\frac{1}{l'(l'+1)}\right)
\end{equation}
Due to Lorentz violation, $R \rightarrow R_0+\delta R$; thus,
$H_{\rm rot} \rightarrow H_{\rm rot}+\delta H_{\rm rot}$, so the
expectation value for the relative shift of the energy levels is
\begin{equation}
\frac{\delta \omega_{\rm rot}(l\rightarrow l')}{\omega_{\rm
rot}(l\rightarrow l')}=\frac{\left<\delta H_{\rm
rot}\right>}{H_{\rm rot}}= -2\frac{\delta R}{R_0}\,,
\end{equation} where $\delta R/R_0$ is given by Eq.
(\ref{deltarh2rot}).

\subsection{Changes due to $\tr(E)$}
The rotation invariant coefficient $\tr(E)$ contained in the
Lorentz-violating $\delta h$ Eq. (\ref{reducedhamil}) can be absorbed into the conventional
Hamiltonian by scaling the mass of the electron:
\begin{equation}
m_e \rightarrow m_e \left(1-\frac 23 \tr(E) \right) \, .
\end{equation}
This leads to a corresponding scaling of the Bohr radius $a_0=4\pi \epsilon_0 \hbar^2/(m_ee^2)$
and the Rydberg constant $R_\infty=\alpha^2m_ec/(2h)$:
\begin{equation}
a_0 \rightarrow a_0 \left(1+\frac 23 \tr(E) \right) \, , \quad R_\infty \rightarrow R_\infty
\left(1-\frac 23\tr(E)\right) \, .
\end{equation}
We expect that quantities of dimension energy will scale like the Rydberg constant; quantities of
the dimension length should scale like the Bohr radius. In fact, the coefficients of $\tr(E)$ in
the change of the ground state energy are close to $-2/3$, while those in the length change are
close to $+2/3$.

The anisotropic effects due to $E_3$ can, however, not be predicted by such simple arguments.

\section{Tests of Lorentz invariance using molecules}\label{tests}

Experiments using molecules may provide interesting new bounds on
some of the Lorentz-violating SME parameters that enter the
Hamiltonian, Eq. (\ref{reducedhamil}). Present bounds on the SME
tensor $c_{\mu\nu}$ are of the order of $10^{-14}$
for the parameter combinations $c_{XY}$ and $c_{XX}-c_{YY}$, and
$10^{-12}$ for $c_{XX}+c_{YY}-2c_{ZZ}$. They were derived \cite{resSME} from experiments \cite{MHBSP,MMalii} based on optical cavities.

Limits on the Lorentz tensors $b_\mu, d_{\mu\nu}, g_{\lambda \mu
\nu}$, and $H_{\mu \nu}$ from present experiments (e.g.,
clock-comparison experiments \cite{KL99}) are sufficiently
stringent so that we can neglect these quantities. The reason is that
these quantities encode spin-dependent effects, that can be
measured to extremely high resolution by monitoring the
frequencies of transitions between Zeeman or Hyperfine levels in
atoms. For example, from such clock-comparison experiments, $B'_j
\lesssim 10^{-24}$ \cite{KL99} ($mB'_j$ is denoted by $\tilde b_J$ in
the literature); from experiments with spin-polarized solids,
components of $B'_j$ are bounded to a few parts in $10^{-25}$
\cite{SPS}. 

Future experiments on Earth and in Space \cite{SpaceTests} are expected to improve the accuracy of many of the spin-dependent terms and also \cite{resSME,OPTIS} of the components of $c_{\mu\nu}$. 

To surpass the present limits on the spin-dependent terms in experiments using molecules, some transition between states having different spins would have to be probed with about 10\,$\mu$Hz
resolution. This is well below the accuracy that is
possible for hydrogen molecules, so we can assume that all Lorentz
violating quantities except $c_{\mu\nu}$ are zero and concentrate
on setting bounds on $c_{\mu\nu}$.

\subsection{Generic experiment}

If the axis of a molecule (or the axes of an ensemble of
molecules) would be aligned and then rotated (e.g., by fixing the
axis in the laboratory, which is subject to Earth's rotation and
orbital motion), $E_3(\theta,\phi)$ will become time dependent.
Thus, the basic principle of Lorentz symmetry tests using
molecules is to measure a property of the molecule, looking for
any changes caused by a rotation of the molecule in space. Since frequencies are the
quantities in physics that can be measured to highest resolution,
the most promising of such experiments are measurements of
transition frequencies using laser spectroscopy methods. 

For simplicity, we will assume that the molecular axis is fixed in the
laboratory. If, instead, the molecule is in an angular momentum
eigenstate, the sensitivity of the experiment gets modified
according to Eq. (\ref{ezzrot}). Either way, however, a method is needed to fix the quantization axis. There is now a broad range of techniques for aligning molecules (see \cite{laseralignment} for a recent review), e.g., using a strong electrostatic field, a hexapole electrostatic field, or polarized laser fields. Most methods need a nonzero electric dipole moment of the molecule. The dipole moment of HD and HD$^+$ is nonzero, but small ($\sim 5.8\times 10^{-4}$D \cite{Tre68} for HD).  Another method of aligning the molecules, based on pulsed laser fields, uses the anisotropic dipole polarizability and hence also works for molecules having zero dipole moment. A detailed discussion of these techniques is beyond the present scope and can be found in the literature \cite{laseralignment}.

\subsection{Possible experiments}

Experiments may use (i) transitions between different electronic energy levels or (ii) (ro-) vibrational transitions within one electronic state, preferably the ground state.

\subsubsection{Electronic transitions}

Electronic transitions in hydrogen moelcules are difficult to measure to extremely high resolution because of the natural linewidth of those transitions, limited by the lifetime of the excited electronic states. Moreover, the transition frequencies generally lie within the uv region of the spectrum, where narrow-band laser radiation is difficult to generate. For example, the transition between the ground state $X^1\Sigma_g^+$ and the excited state $EF^1\Sigma_g^+$ of neutral H$_2$ is at a wavelength of about $\lambda=80$\,nm (corresponding to a frequency of $\sim 3\times 10^{15}$\,Hz). A combination of frequency multiplied $\sim 10$\,ns pulsed and continuous wave lasers can be used for spectroscopy \cite{Gil92}, which results in linewidths of several tens of MHz, or $\sim 3\times 10^{-9}$ of the transition frequency. 

The signal for Lorentz violation would be the differential shift of the ground state energy, Eq. (\ref{deltaU}), and the energy of the excited electronic state, which has to be calculated separately. Such a calculation can also be based on the Born-Oppenheimer method, starting from a linear combination of atomic orbitals like Eq. (\ref{LCAO}) where one or both of the orbitals are excited states of the hydrogen atom. The resulting differential energy shift will be in analogy to Eq. (\ref{deltaU}), with adjusted numerical coefficients. To improve the present limits on $c_{\mu\nu}$, a electronic transition needs to be probed to a resolution of about $10^{-15}$, wich appears unlikely to be reached in the near future. Thus, it is uneccesary to work out those theoretical details here.

\subsubsection{(Ro-) vibrational transitions}

For H$_2$ and H$_2^+$, the rotational and vibrational transitions within one electronic level are dipole--forbidden due to the symmetry of the molecules. They are allowed for HD and HD$^+$, however.  

An experiment using HD$^+$ is currently in preparation \cite{Schiller03,Froe04}. Ionized molecules are used, because they can be cooled using sympathetic cooling by laser cooled atomic ions, eliminating the Doppler broadening of the transitions that occurs at room temperature as a consequence of the thermal motion of molecules in the gas phase. For the spectroscopy in this experiment, a wavelength of $\lambda=1.4\mu$m may be used (among other possibilities), which would drive transitions between rotational sublevels of the $v=1$ and $v=4$ vibrational levels, see Fig. 1 in \cite{Froe04}; continuous wave laser radiation with excellent spectral properties can be generated at this wavelength. The linewidth is thus expected
to be limited by the natural linewidth, of the order of tens of Hz for the dipole-allowed vibrational transitions in the electronic ground state (and below 1\,Hz for pure rotational transitions in
the lowest vibrational state) \cite{Froe04}. The signal for Lorentz violation for this experiment is given by Eq. ({\ref{deltaomvib}) \cite{remark1}; its time dependence is treated below.

The basic advantage of such experiments with molecules rather than cavities is the lower linewidth of the (ro-) vibrational transitions: While a given relative change of the resonance frequency of an optical cavity \cite{resSME} and a ro-vibrational transition in hydrogen translate into approximately equal estimates for the Lorentz-violating quantities, optical cavities presently used have linewidths of tens of kHz \cite{MHBSP,Braxmaier}; the above linewidths of HD$^+$ are three orders of magnitude lower. Thus, molecular experiments have a potentially higher resolution.

\subsection{Time-dependence of the Signal for Lorentz violation}

Since the quantity $E_3$ transforms in a nontrivial way under Lorentz boosts and rotations, Lorentz violation will lead to a characteristic time dependence of transition frequencies in molecules ('signal'). The explicit time-dependence of $E_3$ is calculated in the appendix. To give an example, consider an experiment based on ro-vibrational transitions as described above, where the axes of an ensemble of molecules would be oriented horizontally, pointing south. The signal can then be obtained from Eq. (\ref{statmol}) by inserting $\vartheta=0$:
\begin{eqnarray}
\frac{\delta\om}{\om} & = & 3\alpha \Big[c_{(YZ)}\sin2\chi\sin\oo\To\nonumber\\
& & -c_{(XY)}\cos^2\chi\sin2\oo\To+c_{(XZ)}\sin2\chi\cos\oo\To\nonumber \\
& & -\frac12(c_{XX}-c_{YY})\cos^2\chi\cos2\oo\To\Big] \, ,
\end{eqnarray}
where $\chi$ is the geographical colatitude at which the experiment is performed and $\oo\simeq 2\pi/$23h56min is the sidereal angular frequency of Earth's rotation. $\alpha$ is the numerical factor of $E_3$ in the applicable expression for the Lorentz-violating shift in the quantity measured, i.e., $\alpha=-0.254$ for a vibrational transition in H$_2^+$, as given by Eq. (\ref{deltaomvib}). Thus, if a resolution of $10^{-16\ldots -15}$ in the measurement of $\delta\om/\om$ is achieved, individual bounds on the Lorentz-violating coefficients $c_{(YZ)}, c_{(XZ)}, c_{(YZ)}$, and $c_{XX}-c_{YY}$ of about $10^{-15}$ can be expected. This compares favourably with present cavity experiments \cite{BrilletHall,MHBSP,MMalii,Braxmaier,TTSME,resSME} and is competitive with future space projects \cite{OPTIS}.

In analogy to cavity experiments, the optimum sensitivity would probably be achieved by rotating the setup on a turntable at $\om_t$. The signal, given in the appendix, then consists of Fourier components around $2\om_t$. While such an experiment requires considerably more effort compared to a setup without turntable, it is sensitive to the additional combination $c_{XX}+c_{YY}-2c_{ZZ}$ of Lorentz violating coefficients. Moreover, $\om_t$ can be chosen to match the time scale of the optimum sensitivity of the experiment, which could be much lower than 12\,h. In this case, not only the accuracy of the data acquired during one rotation will be better, but averaging over many rotations will also lead to improved statistics in the same time of measurement.

\section{Discussion, Summary and outlook}

We investigated the properties of the neutral and ionized hydrogen molecule H$_2$ and H$_2^+$
under the influence of Lorentz violation in the electrons' equation of motion. We find
Lorentz-violating changes of the frequencies of electronic and (ro-) vibrational
transitions, and of the bond length. The latter can be compared to the geometry change of solids
calculated in \cite{resSME}. The calculations for molecules and solids use dissimilar models and
approximations. The agreement of the results (in both cases the Lorentz-violating relative length
change is given by the elements of $E$ and $\tilde E$ together with numerical coefficients of the
order of one) is thus interesting and confirms the reliability of both calculations.

The sensitivity of the molecules' properties on the parameter
combinations $\tr(E)$ and $E_3$ allows to derive upper limits on
Lorentz violation: Precision calculations and measurements of the
ground state energy of H$_2$ agree to about $1\times 10^{-7}$,
which allows us to place limits of this order of magnitude on
$\tr(E)$ and $E_3$.

Precision measurements using molecules, e.g., by high resolution spectroscopy of ro-vibrational energy levels within the electronic ground state sympathetically cooled HD$^+$, may lead to individual bounds on the
Lorentz-violating coefficients $c_{(YZ)}, c_{(XZ)}, c_{(YZ)}$, and $c_{XX}-c_{YY}$ of the order of $10^{-15}$ or better. This would compare favorably to the $10^{-14}$ accuracy of the best present cavity experiments and be competitive to future space tests \cite{OPTIS}. Experiments using a turntable will additionally measure the parameter combination $c_{XX}+c_{YY}-2c_{ZZ}$ and should yield a further improvement in resolution by at least one order of magnitude. 

The theory presented here does not include Lorentz violation in the photon sector, but
it can be extended to include it. This would open up the possibility of additional tests of
Lorentz symmetry using molecules. The issue of the independent measurability
of the electron and photon parameters is discussed in \cite{resSME,Bailey04}.

\acknowledgments We would like to thank H. Dehnen, who suggested
to conduct this research to us. H.M. acknowledges a Feodor-Lynen
fellowship by the Alexander von Humboldt-Stiftung. A.S. acknowledges
financial support by the Stifterverband f\"ur die Deutsche
Wissenschaft and the Fonds der chemischen Industrie. C.L. thanks
the german space agency DLR for financial support.

\appendix*

\section{Explicit time-dependence of the signal}

Here, we give in detail the full signal caused by Lorentz
violation in the electrons' equation of motion. We assume that the
axis of the molecule is fixed horizontally and then
rotated, using a turntable at an angular frequency $\ot$ (as
measured in the laboratory on Earth). The rotation axis is fixed
to point vertically. We use a turntable time scale $t_t$ defined
such that $t_t = 0$ at any one instant when the cavity is pointing
in the $x$ direction of the laboratory frame that is defined
below. The signal for experiments where the molecule has a fixed
horizontal orientation that has an angle $\theta$ with the $x$
axis can be obtained by setting $\ot t_t=\theta$, see below.

Limits on the Lorentz-violating quantities of the SME are usually expressed within a sun-centered celestial equatorial reference frame. As defined in \cite{KL99}, it has the $X$ axis pointing towards the vernal equinox (spring point) at 0\,h right ascension and 0$^\circ$ declination, the $Z$ axis pointing towards the
celestial north pole ($90^\circ$ declination) and the $Y$ axis such as to complete the right handed orthogonal dreibein. Earth's equatorial plane lies in the $X-Y$ plane; its orbital plane is tilted by $\eta \simeq 23^\circ$ with respect to the latter. The time scale $T=0$ when the sun passes the spring point, e.g., on march 20, 2001 at 13:31 UT.

We also define a laboratory frame, which has the $x$ axis pointing south, the $y$ axis east, and the $z$ axis
vertically upwards. The laboratory time scale $\To = 0$ when the $y$ and the $Y$ axis coincide.

The signal derivation starts from the symmetrized tensor $c_{(\mu
\nu)}$ given in the sun-centered celestial equatorial reference
frame. The first step would be a Lorentz boost according to
$\beta_\oplus \simeq 10^{-4}$, the velocity of Earth's orbit.
However, because of the symmetry of the molecule, this turns out
to generate no first order contributions to the signal. Taking
into account also the second order is unnecessary
because it is suppressed by a factor $\beta_\oplus^2 \sim 10^{-8}$.

Application of the rotation matrix 
\begin{equation}\label{rotationmatrix}
R = \left( \begin{array}{ccc}
   \cos \chi \cos \oo \To & \cos \chi \sin \oo \To & -\sin \chi \\
   -\sin \oo \To & \cos \oo \To  & 0 \\
   \sin \chi \cos \oo \To & \sin \chi \sin \oo \To & \cos \chi
\end{array} \right) \, ,
\end{equation}
where $\chi$ is the geographical colatitude, and $\oo \simeq 2\pi/$23\,h\,56\,min Earth's rotation angular frequency, leads to the tensor $c_{\mu\nu}$ as expressed within the laboratory frame. Another rotation around the $z$ axis using the rotation matrix
\begin{equation}
R_t = \left( \begin{array}{ccc} \cos \ot t_t & \sin \ot t_t & 0 \\
-\sin \ot t_t & \cos \ot t_t & 0
\\ 0 & 0 & 1 \end{array} \right)
\end{equation}
leads to the quantities within the rotating turntable frame, which are then decomposed according to Eq.
(\ref{hamiltonianparameters}). A last rotation by a right angle aligns the molecules $z$ axis to the $x_t$ axis.

The properties of hydrogen molecules depend on the parameter
combinations $\tr E$ and $E_3=\tr E-3E_{zz}$. Because of the
rotation invariance of the trace, only $E_3$ becomes time
dependent. $E_3$ can be expressed as a Fourier series
\begin{eqnarray}
E_3&=& C(0,0)+ \sum_{a,b} [ S(a,b) \sin (a\ot
t_t+b\oo\To)\nonumber
\\&&+ C(a,b) \cos (a\ot t_t+b\oo\To)]
\end{eqnarray}
with coefficients $S(a,b)$ and $C(a,b)$; the dc component $C(0,0)$
is not included in the equations below, as it is not measurable.
We have seven signal frequencies with the amplitudes
\begin{eqnarray*}
C(0,1) & = & 3c_{(XZ)}\cos\chi\sin\chi \, , \\
S(0,1) & = & 3c_{(YZ)} \cos \chi \sin \chi \, , \\
C(0,2) & = & \frac34 (c_{XX}-c_{YY})\sin^2\chi \, , \\
S(0,2) & = & \frac32 c_{(XY)} \sin^2\chi \, , \\
C(2,-2) & = &-\frac32(c_{XX}-c_{YY})\sin^4\frac\chi2 \, ,
\\ S(2,-2) & = & 3c_{(XY)}\sin^4\frac\chi2 \, ,
\\ C(2,-1) & = & -6 c_{(XZ)}\cos \frac\chi2 \sin^3\frac\chi2 \, , \\ S(2,-1) & = &
6c_{(YZ)}\cos\frac\chi2\sin^3\frac\chi2 \, , \\
C(2,0) & = & \frac34(c_{XX}+c_{YY}-2c_{ZZ})\sin^2\chi \, ,
\\ S(2,0) & = & 0 \, ,
\\C(2,1) & = & 6c_{(XZ)}\cos^3\frac\chi2\sin\frac\chi2 \, , \\ S(2,1) & = & 6c_{(YZ)}
\cos^3\frac\chi2 \sin\frac \chi2 \, , \\ C(2,2) & = &
-\frac32(c_{XX}-c_{YY})\cos^4\frac\chi2 \, , \\
S(2,2) & = & -3c_{(XY)}\cos^4\frac\chi2 \, .
\end{eqnarray*}
The time dependence for experiments without turntable can be
obtained from these coefficients by letting $\omega_t t_t =
\vartheta$, where $\vartheta$ is the angle of the (horizontally
oriented) molecular axis with respect to the north-south axis. We
obtain
\begin{eqnarray}\label{statmol}
E_3  =  \sum_{a,b} [S(a,b)\cos a\vartheta-C(a,b)\sin a\vartheta]
\sin b \omega_\oplus T_\oplus \nonumber
\\ + [S(a,b)\sin a\vartheta +C(a,b) \cos a\vartheta] \cos b \oo \To
\,.
\end{eqnarray}


\begin{references}
\bibitem{missprints} We note misprints in Ref. \cite{fluegge}: On p. 86, two signs have been reversed
in the equation giving $\varphi(\rho)$ such that the energy of
H$_2$ would go to $-\infty$ as $R\rightarrow 0$. The correct
expression reads $\varphi(\rho)=S(\rho)^2(\log
\rho+C)+S(-\rho)^2E_1(4\rho)-2S(\rho)S(-\rho)E_1(2\rho)$.
\bibitem{remark} Here, we used
$\left<p_x^2\right>E_{xx}+\left<p_y^2\right>E_{yy}+\left<p_z^2\right>E_{zz}=\left(\frac
23 \left<p_x^2\right>+\frac 13
\left<p_z^2\right>\right)\tr(E)+\frac 13 \left(\left<p_x^2\right>-
\left<p_z^2\right>\right)E_3$.
\end{references}
\end{document}